\documentclass[12pt]{article}

\usepackage{epsfig}
\begin{document}
\title{Scaled variational computation of the energy spectrum of a two-dimensional 
hydrogenic donor in a 
magnetic field of arbitrary strength}
\author{Ramiro Pino\thanks{e-mail: rpino@ivic.ivic.ve} \,  and  
V\'{\i}ctor M. Villalba\thanks{e-mail: villalba@ivic.ivic.ve} \\
Centro de F\'{\i}sica,\\
Instituto Venezolano de Investigaciones Cient\'{\i}ficas, IVIC\\
Apdo 21827, Caracas 1020-A, Venezuela}

\maketitle

\begin{abstract}
We compute the energy levels of a 2D Hydrogen atom when a constant magnetic
field is applied. With the help of a mixed-basis variational method and a generalization 
of virial theorem, which consists in scaling the wave function, we calculate 
the binding energies of the
1S, $2P^-$ and $3D^-$ levels. We compare the computed energy spectra with
those obtained via a generalization of the mesh point technique as well as
the shifted $1/N$ method. We show that the variational solutions present a
very good behavior in the weak and strong magnetic field regimes.
\end{abstract}
 

In the past few years a great effort has been devoted to the study of  
two-dimensional atoms. The technological 
advances in semiconductor  processing has enabled to manufacture nanostructures
where the electrons are confined to two dimensional (quantum wells), one dimensional 
(quantum wires) and zero dimensional (quantum dots) motions 
(\cite{Bastard,Mendez,Greene0,Greene,Sikorski,Kastner} and references therein). Because of the reduced 
dimensionality, new and interesting transport and optical properties can be 
observed. In order to explain the experimental data, a large number of theoretical 
investigations have been performed\cite{Maksym,Johnson,Proetto}. 

Probably the best known quantum well
configuration consists of regions of GaAs which act as wells of 
conduction electrons, separated by regions of Ga$_{1-x}$Al$_{x}$As which act
as barriers. The application of magnetic field perpendicular to the quantum wells 
is expected to provide further band-structure data and binding energy information. 
The importance of the theoretical calculations demands reliable analytical and 
numerical techniques. In this study we are motivated to analyze the binding 
energies of the ground and excited states of a hydrogenic donor in a 2D electron 
gas. Here, with the help of a scaled mixed variational method,  we compute the 
energy levels of a 2D Hydrogenic atom in the presence of a magnetic field of 
arbitrary strength,we compare our results with those obtained by numerical 
computation and with the shifted $1/N$ method \cite{Imbo1,Imbo2,Mustafa1}.  

The Hamiltonian describing the electromagnetic interaction between a
conduction electron and a donor impurity center when a constant magnetic
field B is applied perpendicular to the x-y plane can be written as

\begin{equation}
H=-\nabla ^{2}+\gamma L_{z}-\frac{2}{\rho }+\frac{\gamma ^{2}\rho ^{2}}{4}
\label{Ham}
\end{equation}
where we have written the vector potential {\bf A} =$\frac{B}{2}(-y,x,0)=%
\frac{Br}{2}\hat{e}_{\varphi }$ in the symmetric gauge. The coupling
constant $\gamma ,$ which measures the ratio between the magnetic energy and
Coulomb energy, is defined as $\gamma =\epsilon ^{2}\hbar
^{3}B/(ce^{3}m^{*2})$ where $m^{*}$ is the effective mass, $\epsilon $ the
dielectric constant of the host material, $\nabla ^{2}$ is the
two-dimensional Laplacian, and $L_{z}$ is the angular momentum operator $-i\hbar
\partial /\partial \phi $ with eigenvalue $\hbar m. $The units of energy are
given in terms of the effective Rydberg constant ${\mathcal R}%
_{0}^{*}=m^{*}e^{4}/2\hbar ^{2}\epsilon ^{2}$ and the effective Bohr radius $%
a^{*}=\hbar ^{2}\epsilon /m^{*}e^{2},$ respectively.

The substitution 
\begin{equation}
\Psi ({\bf r})=e^{im\varphi }\Psi (\rho ),
\end{equation}
reduces the Schr\"{o}dinger equation $H\Psi =E\Psi $ to the following second
order ordinary differential equation 
\begin{equation}
\left[ -\frac{d^{2}}{d\rho ^{2}}-\frac{1}{\rho }\frac{d}{d\rho }+\frac{m^{2}%
}{\rho ^{2}}+\frac{\gamma ^{2}\rho ^{2}}{4}-\frac{2}{\rho }+m\gamma
-E\right] \Psi (\rho )=0.  \label{ecua}
\end{equation}

Exact solutions of eq. (\ref{ecua}) cannot be expressed in closed form in
terms of special functions. There are analytical expressions for the energy
for particular values of $\gamma $ and $m$ \cite{Lozanskii,Taut1,Taut2}. 
Besides numerical
and perturbation methods \cite{Whittaker,Duggan}, different techniques have
been used in order to obtain the eigenvalues $E$ in equation (\ref{ecua}), 
in particular the two-point Pad\'{e} approximation \cite{MacDonald,Martin} 

Recently, using the shifted 1/N expansion, Mustafa \cite{Mustafa1,Mustafa2}
has computed the $1S$, $2P^{-}$ and $3D^{-}$ energy levels for a 2D donor
impurity in the presence of a magnetic field. The shifted $1/N$ and Pad\'e 
methods are very powerful in computing the energy spectra but
fail in giving a reasonable simple wave function and general criteria for
the lower bound of energy values. In the present article we discuss the
problem using a scaled two-terms mixed-basis variational approach. We
compare our results with those obtained using the Schwartz \cite{Schwartz}
interpolation technique, as well as the shifted $1/N$ method; showing that
our results fit very well to those computed numerically for any range of
values of the magnetic strength $B$.

In order to apply the variational method to our problem \cite{Davydov}, we
look for a trial wave function. Since equation (\ref{ecua}) reduces to the
Hydrogen atom equation when $\gamma =0,$ we can consider as a basis for $%
\gamma <<1$ the Hydrogen wave functions $\Psi _{H}.$ Since $<\Psi _{H}\left|
H\right| \Psi _{H}>\geq E$, we obtain a lower bound of the energy for small
values of the parameter $\gamma .$ The solution of equation (\ref{ecua}) when $%
\gamma =0$ is

\begin{equation}
\Psi _{H}(\rho )=D_{m,n}e^{-\rho /(1/2+n_{\rho }+\left| m\right| )}\rho
^{\left| m\right| }L(n_{\rho },2\left| m\right| ,\frac{2\rho }{(1/2+n_{\rho
}+\left| m\right| )})
\end{equation}
where $D_{m,n}$ is a normalization constant, and L(a,b,x) are the Laguerre
polynomials\cite{Lebedev}. Consequently the energy spectrum in the
zero-field limit takes the form 
\begin{equation}
E_H=-\frac 1{(1/2+n_\rho +\left| m\right| )^2}
\end{equation}

Conversely, for large values of $\gamma ,$ a good trial basis is that of the
spherical oscillator. In this case the solution of equation (\ref{ecua}) has the
form 
\begin{equation}
\Psi _{Osc}(r)=C_{m,n}e^{-\gamma \rho ^{2}/4}\rho ^{\left| m\right|
}L(n_{\rho },\left| m\right| ,\frac{\gamma }{2}\rho ^{2})
\end{equation}
and in the high-field limit, the energy levels are 
\begin{equation}
E_{Osc}=\gamma (2n_{\rho }+\left| m\right| +m+1).
\end{equation}

If we attempt to apply the variational method using the hydrogen atom basis,
we will obtain good agreement with the accurate results for small values
of $\gamma ,$ but this approach fails for large $\gamma $ even if we
consider a many term basis. An analogous situation occurs when we use the
oscillator basis, in which case we obtain a good agreement for large $\gamma $
but the convergence is very slow for small values of $\gamma $ \cite
{villalba}. In order to solve this problem, we propose a mixed-basis
approach. The idea is to use as trial function, for any quantum level, a
linear combination of the form 
\begin{equation}
\Psi =c_{H}\Psi _{H}+c_{O}\Psi _{Osc}
\end{equation}
where $\Psi _{H}$ and $\Psi _{Osc}$ are the corresponding hydrogen and
oscillator wave functions associated with the quantum level in question; $c_{O}$
and $c_{H}$ are constants to be calculated. It is worth noticing
that our mixed basis is not orthogonal under the inner product $%
\int_{0}^{\infty }\Psi _{i}\Psi _{j}\rho d\rho .$ We proceed to minimize the
expectation value $\left\langle \Psi \left| H\right| \Psi \right\rangle $
with the normalization condition, $\left\langle \Psi \mid \Psi \right\rangle
=1$

Applying the variational approach to the basis coefficients $c_{i},$ we reduce our
problem to that of solving the matrix equation

\begin{equation}
\left[ \left\langle \Psi _{i}\left| H\right| \Psi _{j}\right\rangle -\lambda
\left\langle \Psi _{i}\mid \Psi _{j}\right\rangle \right] c_{j}=0,
\label{var}
\end{equation}
whence after substituting the Hamiltonian (\ref{Ham}) into (\ref{var}), and
taking into account the separated differential equation (\ref{ecua}) we
obtain the matrix equation, 
\[
\left( 
\begin{array}{ll}
E_{H}+\frac{\gamma ^{2}}{4}A+m\gamma -\lambda & (E_{Osc}-\lambda )S-2C \\ 
(E_{Osc}-\lambda )S-2C & E_{O}-2B-\lambda
\end{array}
\right) \left( 
\begin{array}{l}
c_{1} \\ 
c_{2}
\end{array}
\right) =\left( 
\begin{array}{l}
0 \\ 
0
\end{array}
\right) 
\]
where 
\begin{eqnarray}
\left\langle \Psi _{H}\left| \rho ^{2}\right| \Psi _{H}\right\rangle =A,\
\left\langle \Psi _{Osc}\left| \rho ^{2}\right| \Psi _{H}\right\rangle =D,\
\left\langle \Psi _{Osc}\left| \frac{1}{\rho }\right| \Psi _{H}\right\rangle
=C,\nonumber \\
 \left\langle \Psi _{Osc}\left| \frac{1}{\rho }\right| \Psi
_{Osc}\right\rangle =B,  \left\langle \Psi _{H}|\Psi _{0}\right\rangle =S
\end{eqnarray}
where the lowest value of $\lambda $ will be the energy of the level. Since
we obtain a second order equation for $\lambda ,$ we can analytically
compute the energy eigenvalues and eigenvectors. The advantage of this
approach is twofold. First, because of variational approximation, the eigenvalues 
satisfy the
inequality $\lambda \geq E$ and therefore we have a lower bound for our energy
levels. Second, we obtain a relatively simple expression for the normalized
eigenfunctions.

Instead of restricting ourselves to the the energy values obtained from (\ref
{var}) via the mixed-basis variational method, we can improve the accuracy
of the energy with the help of the virial theorem. Using this technique, after 
re-scaling the radial parameter $r\rightarrow \xi r$ we
have that 
\begin{equation}
\label{scaled}
\Psi (\rho )\rightarrow \Psi (\xi \rho )
\end{equation}
and the matrix terms associated with $\rho ^{n}$ and the second derivative
become 
\begin{equation}
\left\langle \Psi _{i}\left| \rho ^{n}\right| \Psi _{j}\right\rangle
\rightarrow \xi ^{-n}\left\langle \Psi _{i}(\xi r)\left| (\xi \rho
)^{n}\right| \Psi _{j}(\xi \rho )\right\rangle =\xi ^{-n}\left\langle \Psi
_{i}\left| \rho ^{n}\right| \Psi _{j}\right\rangle
\end{equation}
\begin{equation}
\left\langle \Psi _{i}\left| \frac{d}{d\rho ^{2}}\right| \Psi
_{j}\right\rangle \rightarrow \xi ^{2}\left\langle \Psi _{i}(\xi \rho
)\left| \frac{d}{d(\xi \rho )^{2}}\right| \Psi _{j}(\xi \rho )\right\rangle
=\xi ^{2}\left\langle \Psi _{i}\left| \frac{d}{d\rho ^{2}}\right| \Psi
_{j}\right\rangle
\end{equation}
Then, with the help of the virial theorem for a Coulomb and a radial
oscillator potential, 
\begin{equation}
E_{H}=-\left\langle \Psi _{H}\left| T_{H}\right| \Psi _{H}\right\rangle =%
\frac{1}{2}\left\langle \Psi _{H}\left| V_{H}\right| \Psi _{H}\right\rangle
\end{equation}
\begin{equation}
E_{Osc}=2\left\langle \Psi _{Osc}\left| T_{Osc}\right| \Psi
_{Osc}\right\rangle =2\left\langle \Psi _{Osc}\left| V_{Osc}\right| \Psi
_{Osc}\right\rangle
\end{equation}
where $T$ is the kinetic energy term, the re-scaled Hamiltonian
reduces to 
\begin{eqnarray}
H=\left( 
\begin{array}{ll}
H_{11}(\xi ) & H_{12}(\xi ) \\ 
H_{21}(\xi ) & H_{22}(\xi )
\end{array}
\right) =\\ \nonumber
\left( 
\begin{array}{ll}
(\xi (2-\xi )E_{H}+\frac{\gamma ^{2}}{4\xi ^{2}}A+m\gamma & \xi ^{2}J+\frac{%
\gamma ^{2}}{4\xi ^{2}}D-2\xi C+mS \\ 
\xi ^{2}J+\frac{\gamma ^{2}}{4\xi ^{2}}D-2\xi C+mS & \frac{1}{2}(\xi
^{2}-\xi ^{-2})(E_{Osc}-m\gamma )-2\xi B+m\gamma
\end{array}
\right)
\end{eqnarray}
with 
\begin{equation}
J=\int_{0}^{\infty }(\frac{d\Psi _{H}}{d\rho }\frac{d\Psi _{Osc}}{d\rho }%
+m^{2}\Psi _{H}\Psi _{Osc})\rho d\rho.
\end{equation}
In order to compute the quantum energy values, we proceed to compute the
lowest eigenvalue of the matrix equation given by 
\begin{equation}
\left( 
\begin{array}{ll}
H_{11}(\xi )-\lambda & H_{12}(\xi )-\lambda S \\ 
H_{12}(\xi )-\lambda S & H_{22}(\xi )-\lambda
\end{array}
\right) \left( 
\begin{array}{l}
c_{1} \\ 
c_{2}
\end{array}
\right) =\left( 
\begin{array}{l}
0 \\ 
0
\end{array}
\right)
\end{equation}
which takes the form 
\begin{eqnarray}
\lambda =\frac{H_{11}+H_{22}-2SH_{12}}{2(1-S^{2})}\\ \nonumber
-\frac{\sqrt{%
(H_{11}+H_{22}-2SH_{12})^{2}-4(1-S^{2})(H_{11}H_{22}-H_{12}^{2})}}{2(1-S^{2})%
}.
\end{eqnarray}
Since the components of $H$ depend on the scale parameter $\xi $, $\lambda $ is a 
function of $\xi $, and the optimal value for $\xi $
can be obtained computing the minimum for $\lambda (\xi )$, which reduces
to solving the algebraic equation 
\begin{equation}
\label{vari}
\frac{d\lambda }{d\xi }=0.
\end{equation}
Then, we can substitute the value of $\xi$, obtained after solving (\ref{vari}), 
into (\ref{scaled}) and obtain an analytic expression for the scaled wave functions.  
As illustration of the mixed-basis method, we present two different expansions of the 
ground state of th 2D Hydrogen Hamiltonian (\ref{Ham}). We
plot the energy against $\gamma ^{\prime }=\gamma /(\gamma +1)$ as the
horizontal scale. 

For comparison, the numerical computations of the energy spectra associated with equation (\ref
{ecua}) are carried out with the help of the Schwartz method \cite
{Schwartz} which is a generalization of the mesh point technique for
numerical approximation of functions. This method gives highly accurate
results given a thoughtful choice of the reference function. For equation (\ref
{ecua}) we chose as the interpolation function 
\begin{equation}
f(\rho )=\sum_{m}f_{m}\frac{u(\rho )}{(\rho -r_{m})a_{m}},
\end{equation}
where 
\begin{equation}
u(\rho )=\sin [\pi (\rho /h)^{1/2}].
\end{equation}
Here $r_m$ is a zero of $u(\rho )$, $a_m$ is a zero of its derivative, and $h$ is
the step of the quadratic mesh.
Here, as illustration of the mixed-basis method, we present two different
expansions of the ground state of the 2D Hydrogen Hamiltonian 
\begin{figure}
\begin{center}
\epsfig{file=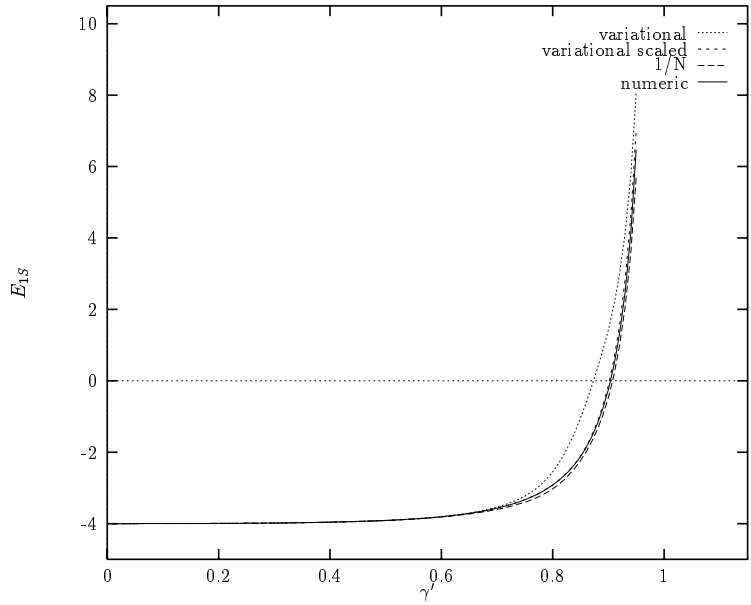,height=8cm}
\end{center}
\caption{Energy of the $1S$ state as a function of $\gamma^{\prime}$. The
thin solid line is obtained by numerical methods; the dotted line is obtained 
using the mixed 1S hydrogen and 1S oscillator bases variational method. The 
thin broken  line is obtained via the scaled variational method with the 1S 
hydrogen and oscillator bases. The thick broken line is obtained with the help 
of the shifted $1/N$ method}
\end{figure}
\begin{figure}
\begin{center}
\epsfig{file=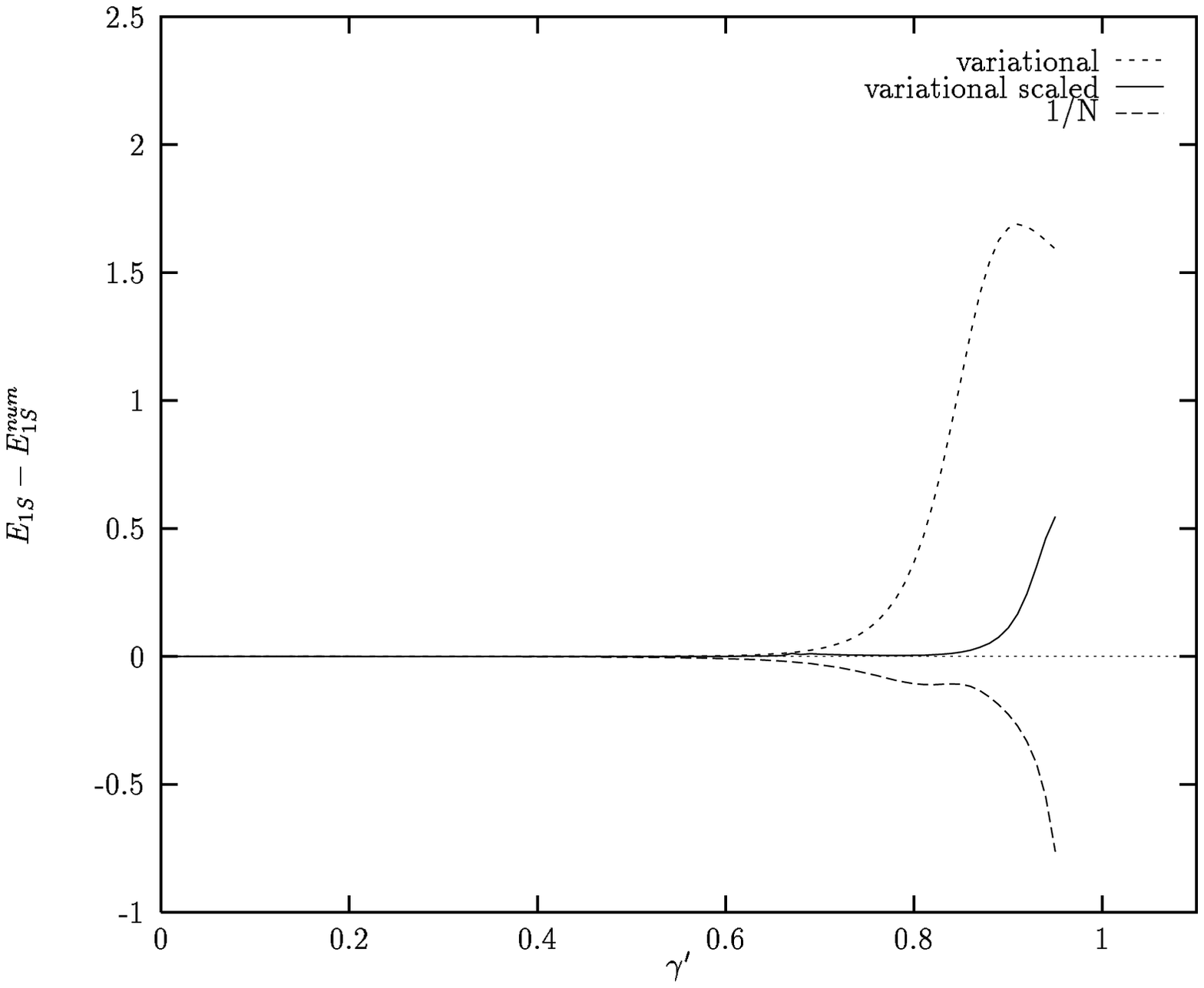,height=8cm}  
\end{center}
\caption{The figure shows the difference between the numerical result for the 
$1S$ energy spectrum and the energy values computed with the help of the mixed 
variational (thin broken line), scaled variational (thin solid line), and the 
shifted $1/N$ method (thick broken line)}
\end{figure}

\begin{figure}
\begin{center}
\epsfig{file=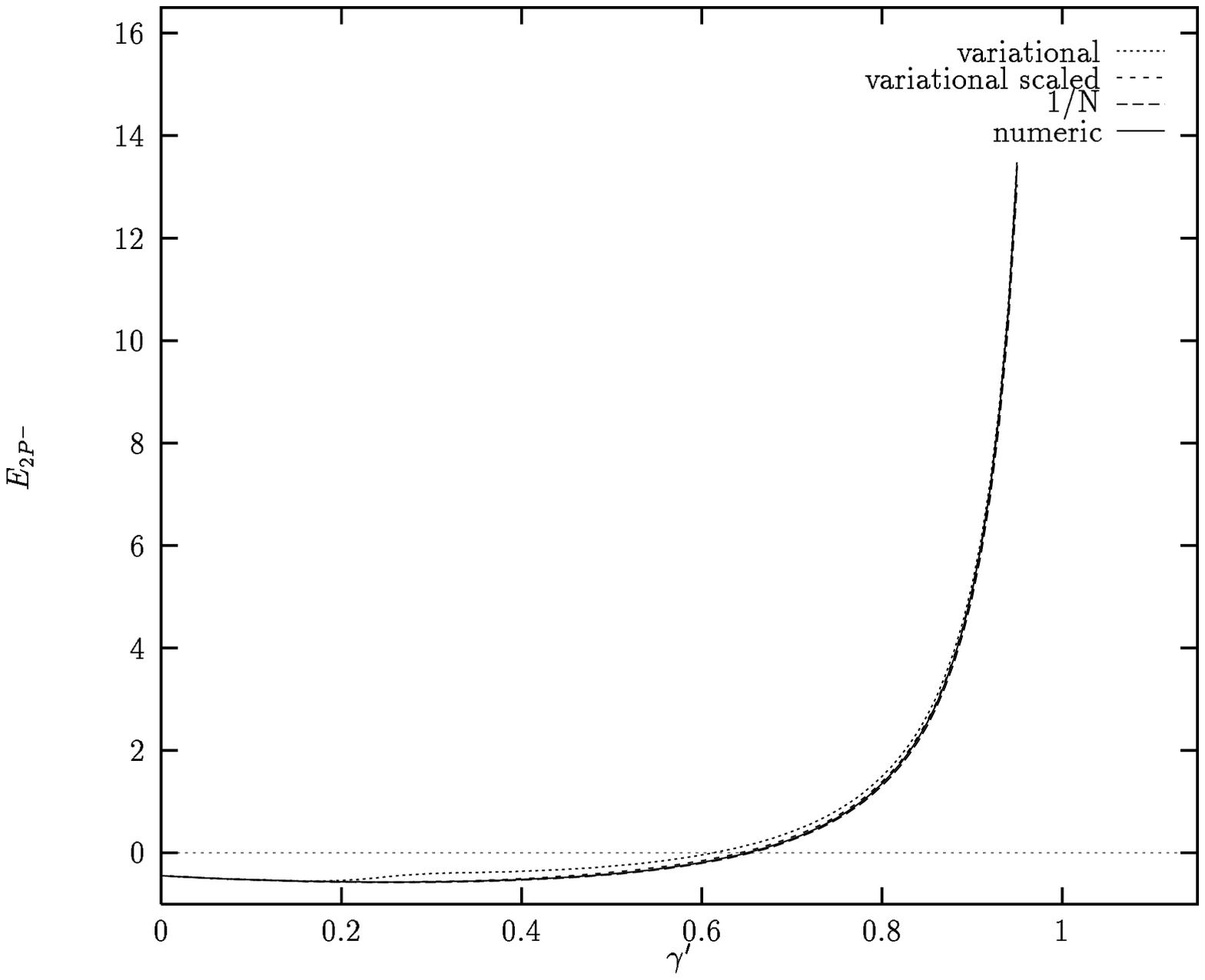,height=8cm}
\end{center}
\caption{Energy of the $2P^-$ state as a function of $\gamma^{\prime}$. The
thin solid line is obtained by numerical methods; the dotted line is obtained 
using the mixed $2P^-$ hydrogen and $2P^-$ oscillator bases variational method. 
The thin broken line is obtained via the scaled variational method with the $2P^-$ 
hydrogen and oscillator bases. The thick broken line is obtained with the help of 
the shifted $1/N$ method} 
\end{figure}
\begin{figure}
\begin{center}
\epsfig{file=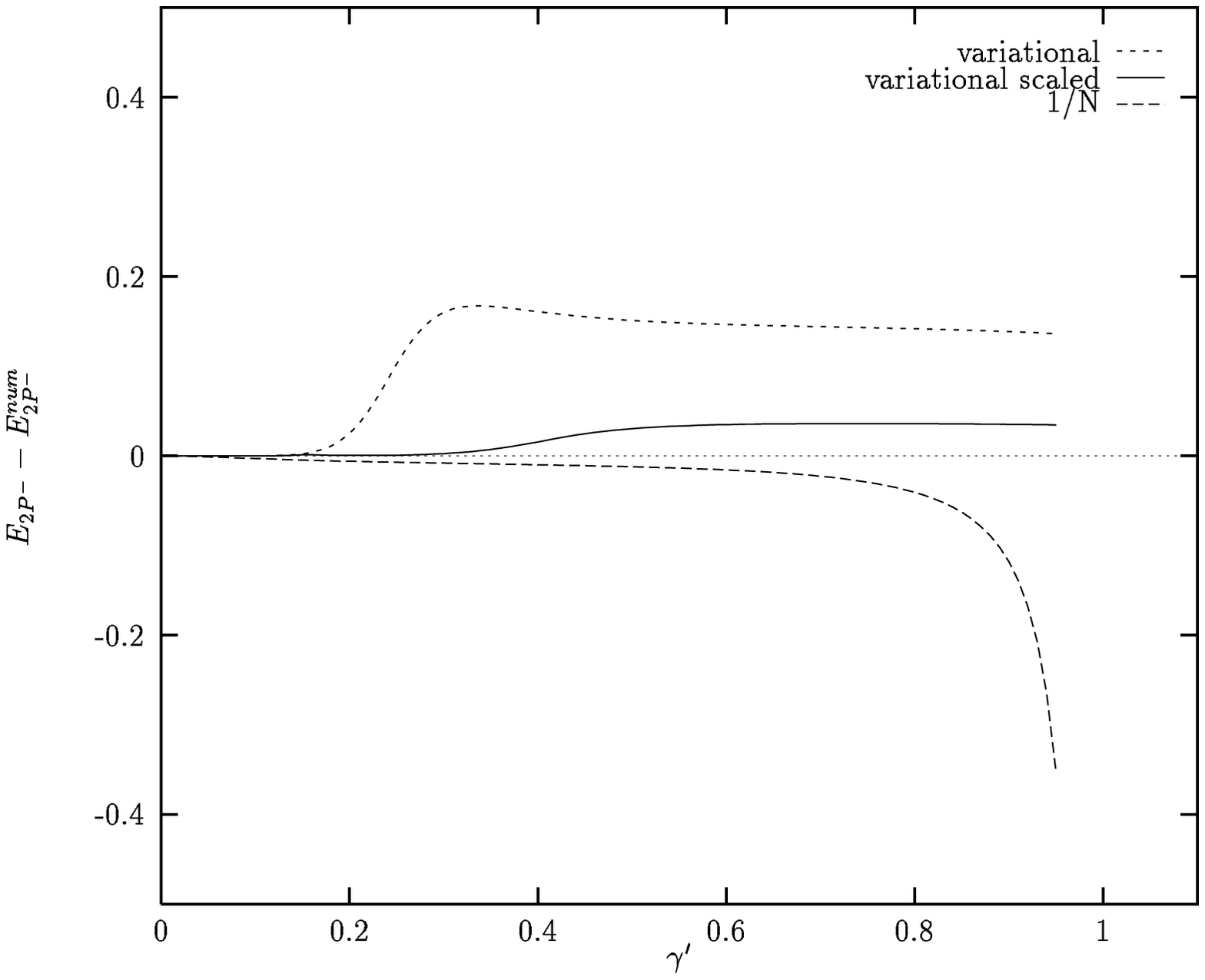,height=8cm}
\end{center}
\caption{The figure shows the difference between the numerical result for the 
$2P^-$ energy spectrum and the energy values computed with the help of the mixed 
variational (thin broken line), scaled variational (thin solid line), and the 
shifted $1/N$ method (thick broken line)}
\end{figure}
\begin{figure}
\begin{center}
\epsfig{file=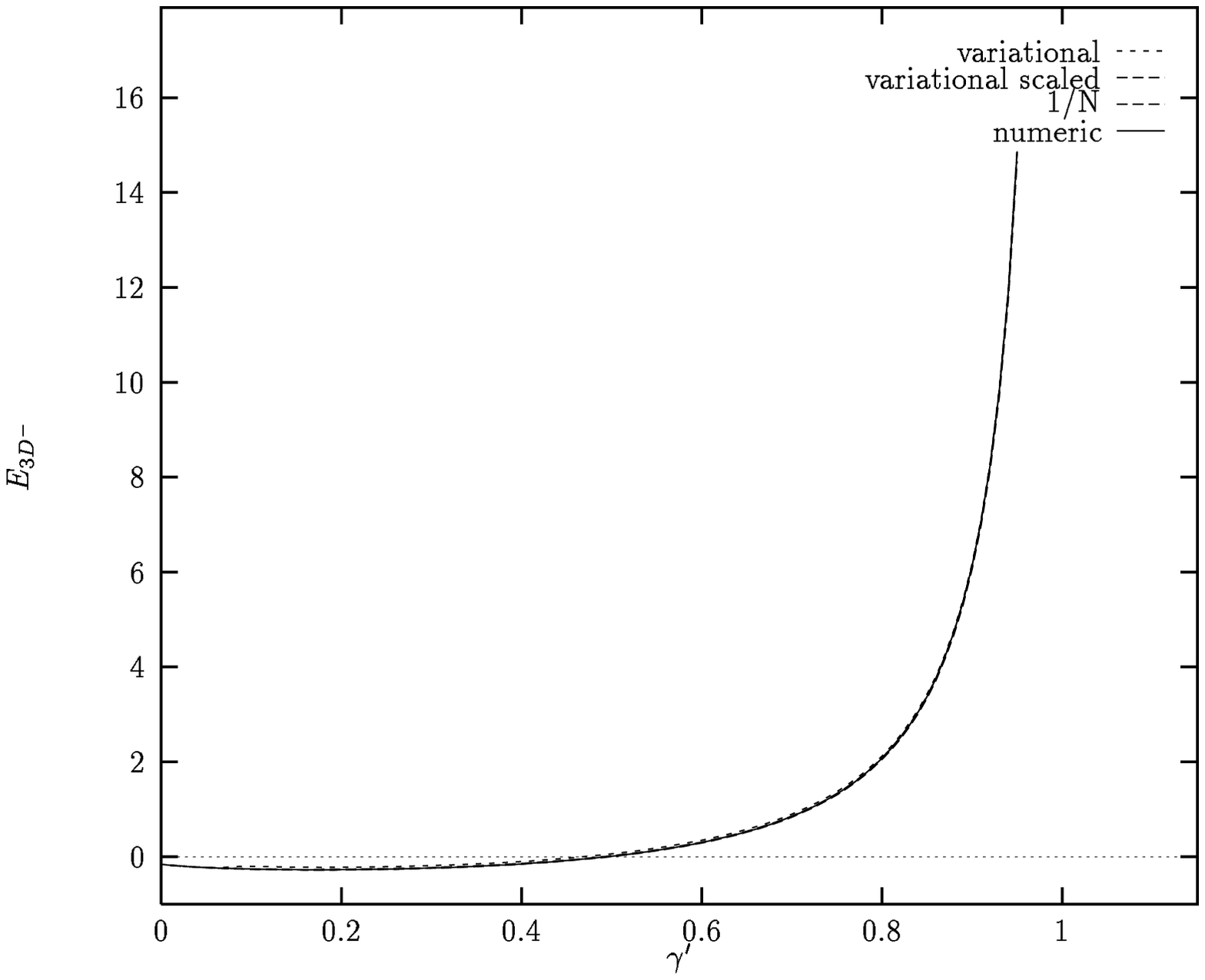,height=8cm}
\end{center}
\caption{Energy of the $3D^-$ state as a function of $\gamma^{\prime}$. The
thin solid line is obtained by numerical methods; the dotted line is obtained 
using the mixed $3D^-$ hydrogen and $3D^-$ oscillator bases variational method. 
The thin broken line is obtained via the scaled variational method with the $3D^-$ 
hydrogen and oscillator bases. The thick broken line is obtained with the help of the 
shifted $1/N$ method} 
\end{figure}
\begin{figure}
\begin{center}
\epsfig{file=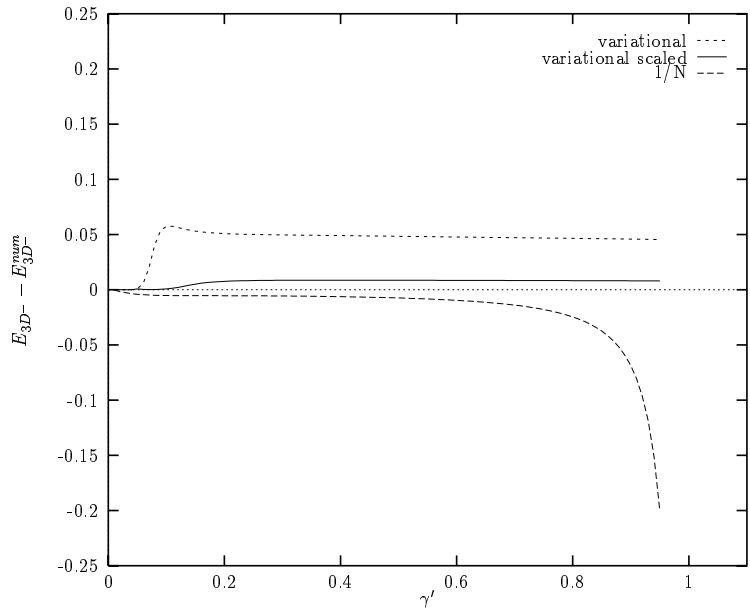,height=8cm}
\end{center}
\caption{The figure shows the difference between the numeric result for the $3D^-$ 
energy spectrum and the energy values computed with the help of the mixed variational (thin broken line), scaled variational (thin solid line), and the shifted $1/N$ method (thick broken line)}
\end{figure}
Figures 1, 3, and 5  compare the variational results
with those obtained numerically and with the help of the shifted $1/N$ method.

It is easy to see that even for a two term mixed basis a good fitting is
obtained in the weak and strong field regimes. One of the bases gives a
reasonable good fitting in the intermediate region. A better fit is 
obtained with the help of the scaled variational method. We also have
that for the $2P^-$ and $3D^-$ states the mixed-basis variational approach
gives very good results.

Examining Figs. 1, 3 and 5, it is not obvious which technique gives the most accurate 
results  for computing the hydrogen energy levels. However, Figures 2, 4 and 6 
show that the shifted $1/N$ method always gives results below the the numerical 
energy levels. Among the three analytic methods, figures 2, 4 and 6 show 
that the scaled variational method gives the most close to numerical results 
even for large values of $\gamma$.  

It would be interesting to apply the mixed-basis technique as well as the scaled 
variational method  for the 2D Hydrogen problem when relativistic effects are not negligible. This will be
the object of a forthcoming publication.

\section*{Acknowledgments} We thank Dr. Juan Rivero for helpful discussions. 
This work was supported by CONICIT under project 96000061.

\end{document}